\documentclass[prl,twocolumn, superscriptaddress]{revtex4}

\newif\ifpdf
\ifx\pdfoutput\undefined
\pdffalse % we are not running PDFLaTeX
\else
\pdfoutput=1 % we are running PDFLaTeX
\pdftrue
\fi

\ifpdf
\usepackage[pdftex]{graphicx}
\else
\usepackage{graphicx}
\fi
\usepackage{amsmath}
\usepackage{amssymb}
\usepackage{amstext}
\usepackage{amsxtra}
\usepackage{version}
\ifpdf
\DeclareGraphicsExtensions{.pdf, .jpg}
\else
\DeclareGraphicsExtensions{.eps, .jpg}
\fi
\begin{document}
%Janes commands 
%
\newcommand{\Tc }{T$_{\mathrm C}$\space}
\newcommand{\Tco }{T$_{\mathrm CO}$\space}
\newcommand{\sinth}{\mbox{$\sin\theta/\lambda$}}
\newcommand{\inA}{\mbox{\AA$^{-1}$}}
\newcommand{\mub}{\mbox{$\mu_{B}$}}
\newcommand{\mns}{$-$}
\newcommand{\ddd}{3\textit{d}}
\newcommand{\pp}{2\textit{p}}
\newlength{\minusspace}
\settowidth{\minusspace}{$-$}
\newcommand{\msp}{\hspace*{\minusspace}}
\newlength{\zerospace}
\settowidth{\zerospace}{$0$}
\newcommand{\zsp}{\hspace*{\zerospace}}
\newcommand{\lasrmn}{La$_{2-2x}$Sr$_{1+2x}$Mn$_{2}$O$_7$\space}
\newcommand{\mnt}{$Mn^{3+}$}
\newcommand{\mnf}{$Mn^{4+}$}
\newcommand{\eg}{$\textit{e}_{g}$ }
\newcommand{\qce}{$\textit{q}_{CE}=(1/4,1/4,0)$}
\newcommand{\qcel}{$\textit{q}_{CE}=(1/4,-1/4,0)$}
\newcommand{\qos}{$\textit{q}_{L}=(0.3,0,1)$}
\newcommand{\sqw}{S(\textit{Q},$\omega$)}
\newcommand{\Ts}{T$^{*}$\space}
\newcommand{\GG}{$\Gamma$\space}

%\preprint{}

\title{Glass Transition in the Polaron Dynamics of CMR Manganites}
% Optional argument for running titles on pages
%\title[]{}

% repeat the \author .. \affiliation  e\Tc . as needed

% other informatio
\author{D. N. Argyriou}
%\thanks{}
\affiliation{Materials Science Division, Argonne National Laboratory, Argonne, IL, 60439.}
\altaffiliation{Hahn-Meitner-Institut, Glienicker str. 100, Berlin D-14109, Germany}
\email{argyriou@hmi.de}
\author{J.W. Lynn}
\affiliation{NIST Center for Neutron Research, NIST, Gaithersburg, MD 20899}
\author{R. Osborn}
\affiliation{Materials Science Division, Argonne National Laboratory, Argonne, IL, 60439.}
\author{B. Campbell}
\affiliation{Materials Science Division, Argonne National Laboratory, Argonne, IL, 60439.}
\author{J. F. Mitchell}
\affiliation{Materials Science Division, Argonne National Laboratory, Argonne, IL, 60439.}
\author{U. Ruett}
\affiliation{Materials Science Division, Argonne National Laboratory, Argonne, IL, 60439.}
\author{H. N. Bordallo}
\affiliation{Intense Pulsed Neutron Source, Argonne National Laboratory, Argonne, IL, 60439.}
\author{A. Wildes}
\affiliation{Institut Max Von Laue-Paul Langevin, Boite Postale 156, 38042 
Grenoble Cedex 09, France.}
\author{C. D. Ling}
\affiliation{Materials Science Division, Argonne National Laboratory, Argonne, IL, 60439.}

%option in \documen\Tc lass). \noaffiliation is required (may also be

\date{\today}
\begin{abstract}

Neutron scattering measurements on a bilayer manganite near optimal doping show that the short-range polarons correlations are completely dynamic at high T, but then  \textit{freeze} upon cooling to a temperature \Ts$\approx$ 310 K.  This glass transition suggests that the paramagnetic/insulating state arises from an inherent orbital frustration that inhibits the formation of a long range orbital- and charge-ordered state.  Upon further cooling into the ferromagnetic-metallic state (\Tc=114 K), where the polarons melt, the diffuse scattering quickly develops into a propagating, transverse optic phonon.
\end{abstract}
%\pacs{}
%\keywords{}

\maketitle

The role of local charge-correlations and their competition with magnetic and electronic ground states is one of the most striking new features of the transition metal oxides such as high temperature cuprate superconductors and nickelates, but is most elegantly highlighted for the colossal magnetoresistive (CMR) manganites.  Here the double-exchange (DE) interaction competes with an enhanced electron-phonon coupling via the Jahn-Teller active \mnt  ion.  At optimal doping this competition produces a transition from the ferromagnetic-metallic ground state to an intrinsically inhomogeneous paramagnetic-insulating state at elevated temperatures, which is often described as polaronic.\cite{Millis, Roder} However, the high \mnt concentration at these dopings makes correlations between polarons unavoidable. \cite{Adams,Dai,Doloc} Our inelastic neutron scattering results reveal that within the insulating state there is a freezing of dynamic polaron correlations akin to a glass transition at \Ts=310 K, below which a purely elastic component of the scattering is observed.  The discovery of this transition from a \textit{polaron glass} to a \textit{polaron liquid} at \Ts demonstrates the general role that charge and orbital ordering plays in the manganites, and we believe that it is also relevant in the cuprate and nickelate oxides.

 The polaronic state is a result of the strong coupling of charge-orbital and spin degrees of freedom found in the manganites. In undoped manganese perovskites (such as LaMnO$_{3}$) each \mnt O$_{6}$ octahedron contains four \textit{d} electrons on the Mn ion, three $t_{2g}$ electrons that are HundÕs exchange coupled to form a S=3/2 core spin, and one additional electron in a higher-energy doubly-degenerate \eg orbital, giving S=2. The energy of the occupied \eg level can then be lowered by distorting the octahedron and removing this degeneracy.  This Jahn-Teller (JT) distortion couples the magnetic, electron orbital, and lattice degrees of freedom in a natural way.  The distorted JT octahedra can be packed together to form a long-range orbitally ordered lattice, that also orders antiferromagneticaly as in LaMnO$_{3}$.\cite{cjteffect} The introduction of divalent ions such as Ca$^{2+}$ for La$^{3+}$ removes an equal number of \eg electrons to introduce JT inactive \mnf  ions and undistorted \mnf O$_{6}$ octahedra.  At half doping the equal numbers of \mnt  and \mnf  combine this charge and orbital ordering, along with antiferromagnetic ordering, to form a single \textit{long-range} ordered phase.  Goodenough predicted that this phase, known as CE,\cite{goodenough} exhibits an ordering of \mnt\space  \ddd$_{3z^{2}-r^{2}}$ orbitals on alternate Mn-sites, and superlattice reflections from these cooperative JT-distortions have been experimentally observed in both cubic and bilayer perovskites (e.g.  La$_{0.5}$Ca$_{0.5}$MnO$_{3}$\cite{Radaelli2, Huang} and LaSr$_{2}$Mn$_{2}$O$_{7}$\cite {Argyriou2}). 

For intermediate doping levels, on the other hand, the orbital ordering generally becomes frustrated, and the double-exchange interaction mediated by hopping of the \eg electrons can then dominate the energetics to produce a regime where the ground state is metallic and ferromagnetic.  Interestingly, it has been found recently that as the system transforms from this ferromagnetic/metallic state to the paramagnetic state, polarons form that trap the \eg electrons and render it insulating.  This polaronic state is intrinsically inhomogeneous, consisting of short-range (10-20 \AA ) charge/orbital correlations and longer range lattice deformations that arise from Jahn-Teller defects.\cite{Doloc, Argyriou,Shimo, Adams} The CMR effect then originates when a magnetic field is applied, producing a spin alignment that favors double exchange and melts the polarons and thereby drives this insulator-to-metal transition.

To investigate the dynamics of polarons we have chosen the \lasrmn layered perovskite manganite, due to the availability of large single crystals that allows for the collection of high quality inelastic scattering data.  There is no twinning possible in the tetragonal crystal structure of these materials (I4/\textit{mmm}, \textit{a}=3.86 \AA\space  and \textit{c}=19.9 \AA\space  at room temperature) which simplifies the interpretation of data. These materials consist of a double perovskite layer of corner-shared MnO$_{6}$ octahedra separated by a simple (La,Sr)O rock salt layer.  The doping level of x=0.38 for our single crystal is near the optimal doping for CMR. In this crystal and in the x=0.4 crystal described in ref. \onlinecite{Doloc} and \onlinecite{campbell}, static short-range charge correlations have been observed in the paramagnetic state with propagation vector \qos, and can be described in terms of the ordering of \ddd$_{3z^{2}-r^{2}}$ orbitals within the $ab$-plane that induce a longitudinal distortion to the lattice.\cite{campbell} The intensity of the diffuse scattering at $q_{L}$ scales directly with the resistivity of the material, and rapidly disappears in the metallic state below the ferromagnetic transition temperature \Tc =114 K. This scaling behavior is similar for CE-type correlations in the CMR pseudo-cubic perovskites, which occur at \qce.\cite{Adams,Dai} However, an opposite behavior has been observed for the CE-correlations in our x=0.38 and 0.4 bilayer manganites using high energy synchrotron x-ray scattering, namely that on cooling from 300 K their intensity decreases as T$\rightarrow$\Tc, suggesting that the scattering is dynamic in origin. \footnote{We confirmed that the diffuse scattering at \textit{q$_{CE}$} and  \textit{q$_{L}$} coexist in these manganites by mapping out the intensity of the scattering via individual scans as a function of position from a x=0.4 crystal (from reference \onlinecite{Doloc}) using a 115KeV, 0.3 mm wide beam at the Advanced Photon Source (BESSRC 11ID-C). Together with long range CE ordering observed at x=0.5 we find that CE-correlations are present in these materials over a wide range of x.} This behavior encouraged us to investigate the inelastic neutron scattering from these CE-correlations over a wide temperature range from 7 K to 460 K.

\begin{figure}[b]
 \includegraphics[scale=0.42]{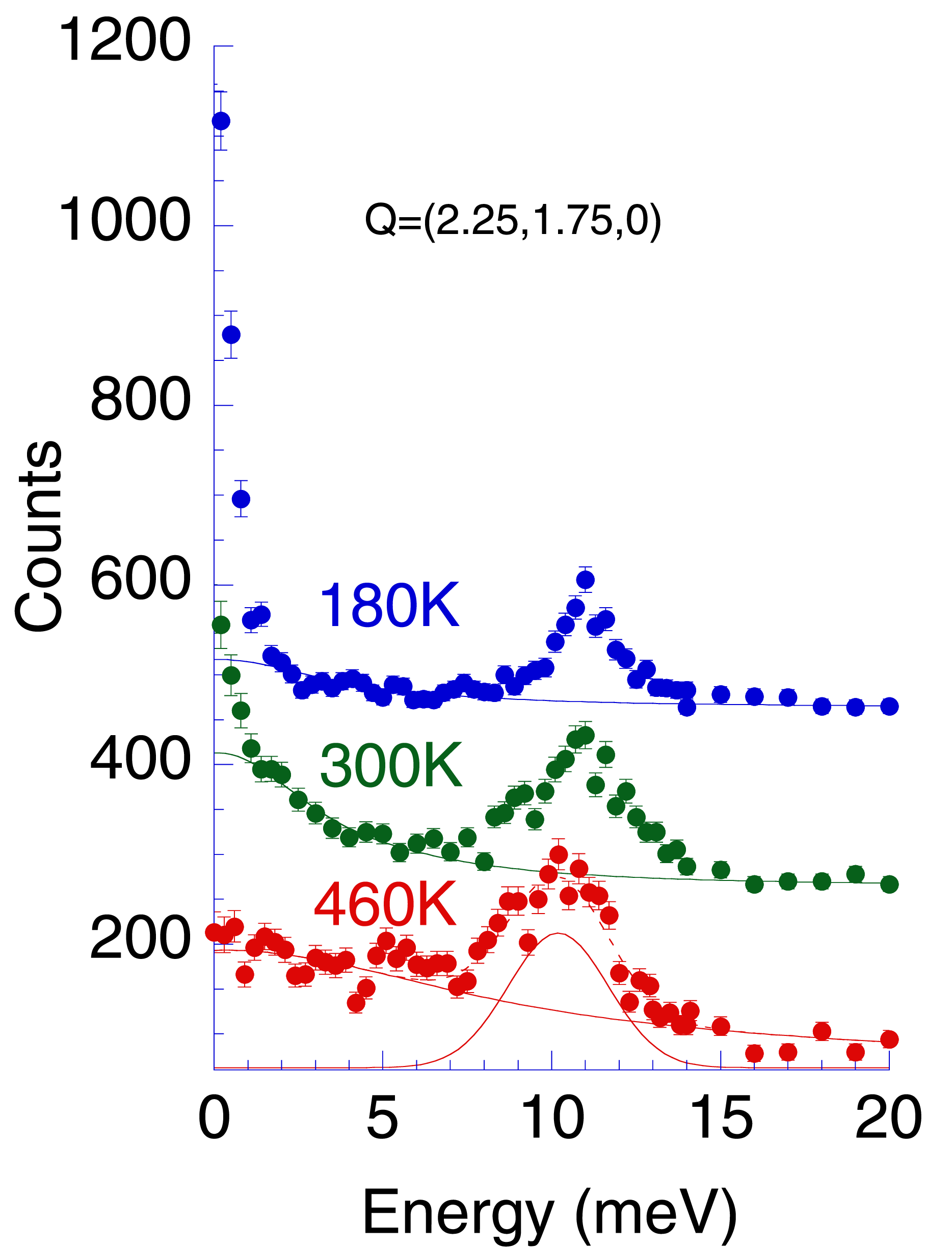}
 \caption{\scriptsize Scans of the observed quasielastic scattering at various temperatures, measured at \textit{Q}=(2.25,1.75,0). The scans measured at 300K and 180K are displaced by 200 and 400 counts respectively for clarity.  Solid lines represent the Lorentzian component, while the Guassian centered at $\approx$11meV describes $[1\overline10]$ transverse acoustic phonon (460K data only).}
 \label{fig1}
 \end{figure}

Inelastic neutron scattering from a single crystal of \lasrmn, x= 0.38 (5mm in diameter and 4cm in height), prepared via the traveling floating zone method, were measured on the BT-2 thermal triple axis spectrometer at the NIST Center for Neuron Research. 60\'-20\'-40\'-80\' \space  collimation and a fixed final energy of 14.8 meV were used. Measurements were made as a function of temperature using a cryo-furnace between 7 to 460 K. The elastic incoherent scattering was measured at 7 K and subtracted from the data taking into account its variation with temperature using Debye-Waller factors from refs.  \onlinecite{Argyriou} and \onlinecite{Mitchell}. The data shown in fig. ~\ref{fig1} were fit using a Gaussian and a Lorentzian response  for the elastic and quasielastic scattering, respectively.  A Gaussian response was also used to account for the transverse acoustic phonon at $\approx$ 11 meV. The various responses were convoluted with the resolution function of the spectrometer and included detailed balance. The results of these fits are shown in fig. ~\ref{fig3}. 

\begin{figure}[!htbp]
 \includegraphics[scale=0.6]{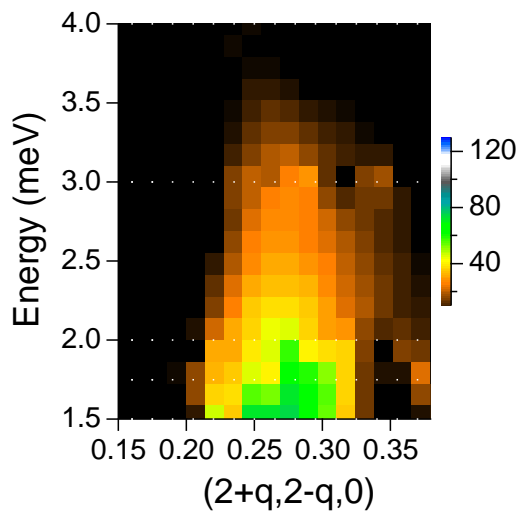}
 \caption{\scriptsize  Color contour map of the quasielastic scattering at 360K. }
 \label{fig2}
 \end{figure}

\begin{figure}[!htbp]
 \includegraphics[scale=0.5]{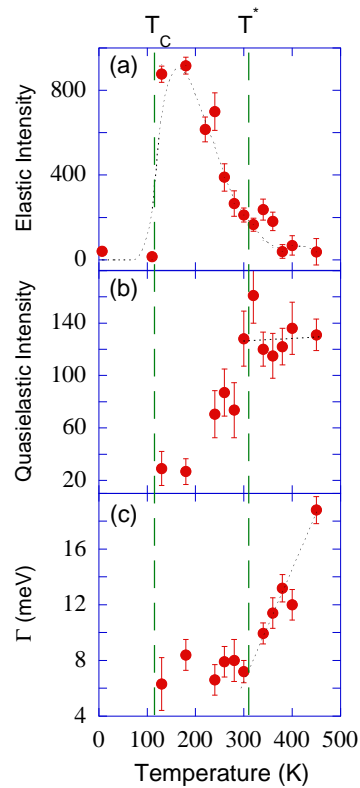}
 \caption{\scriptsize Temperature dependence of the elastic (a) and quasielastic intensity (b) from fitting a Gaussian and Lorentzian response to the spectra shown in fig.  1a measured at \textit{Q}=(2.25,1.75,0). (c) The quasielastic width, $\Gamma$ of the Lorentzian response.  Dashed lines are guides to the eye. Error bars correspond to standard deviations obtained from the least squares analysis.  } 
 \label{fig3}
 \end{figure}

Fig.~\ref{fig1} shows measurements for the CE-type scattering at (2+1/4,2-1/4,0), around the (2,2,0) Bragg reflection.\footnote{This peak has a weak dependence on composition, both in the cubic and layered systems.  For this composition the maximum is observed at (2.27,1.73,0). Given the breadth of the scattering of these short-range correlations, the shift away from the (1/4,1/4,0) position is not significant.} At the highest temperatures the scattering can be well described as \textit{quasielastic}.  The energy width $\Gamma$ of the quasielastic obtained from our fits indicates that the Jahn-Teller polarons fluctuate with a lifetime of the order of femtoseconds ($\tau \sim \hbar
/\Gamma \sim $ 60fs at 360K). To probe further the origin of this quasielastic scattering we performed constant energy scans along the $[1\overline10]$ from the (220) reflection to compile the map shown in fig. ~\ref{fig2}. The transverse phonon and Huang contributions were subtracted to reveal an excess quasielastic scattering centered at \textit{Q}$\sim$ (2.27,1.73,0). This map of the inelastic response shows that the dynamical scattering still has surprisingly well-defined correlations at these elevated temperature, demonstrating that the polarons are not isolated Jahn-Teller \mnt deformations, but are dynamically correlated.  The dynamical correlation length obtained from these data is $\sim$12 \AA , comparable in size to the static correlation length found at lower T. 

These dynamic CE-correlations are still present up to 460 K, the highest T we explored.  On decreasing the temperature from 460 K we see from fig.~\ref{fig1} that the quasielastic scattering narrows in energy, while the intensity of the elastic scattering increases, as shown in the temperature dependence of the fitted intensities and Lorentzian linewidth $\Gamma$ in fig.~\ref{fig3} . At high temperatures the width decreases approximately linearly until T$\equiv$\Ts$\sim$310 K, while the integrated intensity of the scattering is approximately temperature independent.  Below $\sim$310 K, the behavior of the quasielastic width and intensity abruptly changes, and we find that a new, purely elastic component, develops in the spectrum.  The temperature dependence of the fits for both components is also given in fig.~\ref{fig3}.  The data reveal that at high temperatures the neutron spectra are dominated by quasielastic scattering, while on cooling the intensity of the elastic scattering develops rapidly below \Ts$\sim$310 K, and appears to track the resistivity of the material. This increase is accompanied by a concomitant decrease of the quasielastic intensity, while the energy width $\Gamma$ exhibits a change in slope, becoming approximately constant for \Tc $<$T$<$\Ts. In this region the elastic scattering dominates while the quasielastic scattering intensity is non-zero and has a measurable value of $\Gamma$, indicating a coexistence of both frozen and dynamic CE-correlations.  Finally, although our measurements concentrated on the CE-correlations, we found that the quasielastic width of the Huang scattering exhibits similar behavior to that found for the CE-correlations.\footnote{Polaronic correlations at \qos exhibit the same temperature dependence as reported in ref. \cite{Doloc}. Inelastic scans showed that they are mainly elastic in character with an onset at \Ts, in contrast to the scattering at \qcel\space which exhibits a substantial inelastic component as described.}

From our measurements we can identify three regimes.  For T$>$\Ts the observed scattering is dominated by CE-correlations that are completely dynamic.  The quasielastic width from these correlations varies linearly for T$>$\Ts, which is qualitatively what is expected for a continuous phase transition.  We note here that these dynamic correlations have a classical signature in that $k_{B}$\Ts$>>\Gamma$ for the temperature range investigated.  The second regime occurs for \Tc $<$T$<$\Ts, and is dominated by frozen static polaronic correlations as revealed by the rapid development of the CE-elastic component at \Ts. Here the development of the elastic component is accompanied by a decrease of the quasielastic intensity, while the short-range spatial correlations do not diverge.  This behavior is clear evidence for a freezing transition, analogous to the freezing transition T$_{g}$ found in structural glasses, or geometrically frustrated spin-glasses.  For example, in the geometrically frustrated antiferromagnet Y$_{2}$Mo$_{2}$O$_{7}$ a similar transition to a short range ordered magnetic state is observed, with a non-divergent spatial coherence length, development of an elastic component, and a second-order-like behavior of the spin relaxation rate for T$_{g}$. \cite{Gardner} We note that at the same temperature as \Ts, neutron powder diffraction measurements in this bilayer system have identified a strong anomaly in the lattice constants\cite{Mitchell}, and specific heat anomaly.\cite{Gordon} Taken together with the present results the data strongly suggest that \Ts represents a phase transition from a \textit{polaronic liquid} to a \textit{polaronic glass}. This polaron glass then dissolves at \Tc  as the ferromagnetic metallic state sets in. This behavior is in sharp contrast to similar measurements we have conducted in (insulating) LaSr$_{2}$Mn$_{2}$O$_{7}$, that exhibits a long range CE charge ordering transition at 220K. Here critical scattering is observed, with the correlation length of the CE- diffuse peaks diverging at the charge ordering transition, \Tco=210K.

\begin{figure}[!htbp]
 \includegraphics[scale=0.6]{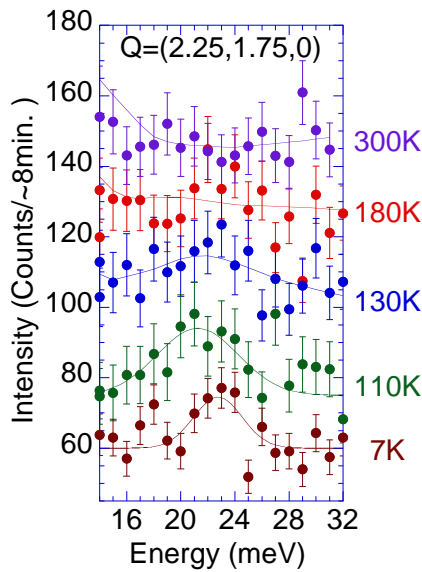}
 \caption{\scriptsize Phonon scans at Q=(2.25,1.75,0) at various temperatures.  An optical phonon is visible at 110K and 7K below \Tc at 21.2(1) and 22.5 meV, respectively.  The same phonon disappears for T$>$\Tc.}
 \label{fig4}
 \end{figure}

The formation of short-range polaron correlations in the paramagnetic state should also strongly affect the lattice dynamics associated with these atomic displacements, and this is indeed the case.  Fig.~\ref{fig4} shows that a well defined transverse optic phonon is observed at the CE peak position at low T (7 K). At 110 K, just below \Tc , the phonon has softened somewhat and broadened, while in the paramagnetic regime (T$>$\Tc  ) the scattering becomes overdamped, evolving into the quasielastic scattering discussed above.  This is similar to the zone folding behavior found recently for the phonons in the high \Tc  cuprates\cite{McQueeney}, associated with the formation of short-range charge stripes, or more general phonon dampening.\cite{Zhang}

The new insight given by our work is the transition at \Ts, which signifies the freezing of dynamic polaron correlations from a polaronic fluid. We demonstrate here that these correlations are indeed dynamic at sufficiently high temperatures, while an unusual insulating state develops below \Ts that is best described as an orbital/charge glass both in terms of its glassy statics and dynamics.  Our results are in tentative agreement with recent quantum MC modeling that predicts a freezing of short range correlations in manganites.\cite{burgy} Hole doping destroys the cooperative JT ordering found in undoped compounds, but it does not remove the JT coupling to the lattice.  The glassy nature of the transition to a inhomogeneous charge state we believe is common to transition metal oxides that exhibit competing ground states.  For optimally doped CMR manganites, the lattice is unable to accommodate long range cooperative ordering of the \eg orbitals and thus is inherently frustrated.  The general behavior of the manganites has parallels in both the doped nickelates and cuprates, where charge order in the form of static or dynamic stripes originates from the same type of competing interactions and frustrations.  The behavior of La$_{1.6-x}$Nd$_{0.4}$Sr$_{x}$CuO$_{4}$, for example, has striking similarities with the manganite that we have examined here (competition between charge stripes and superconductivity as opposed to CE-clusters and double-exchange)\cite{Tranquada}, indicating a commonality of the underlying physics in these \ddd\space transition metal oxides where spin, orbital and lattice degrees of freedom are intimately intertwined. Indeed, the lattice in the cuprates appears to be much more strongly coupled than originally thought\cite{lanzara}, illustrating the universal nature of the underlying physics of the \ddd\space transition metal oxides.  The glass-like behavior of the diffuse scattering and the phonon damping (or zone folding) we observe in the manganites are indications of the frustration that naturally occurs when orbitally-ordered systems are randomized by doping.\cite{Millis2} The orbital frustration inhibits the development of spatial order, but the transition still proceeds as a freezing of the local order, producing a (static) glass.  For the manganites the role of frustration is critical, as it allows for a delicate balance of competing interactions that gives rise to enhanced physical properties such as CMR.

\begin{acknowledgments}
The authors thank K. Gray, D. Komskii, S.-H. Lee, A. Millis, E. Dagatto and P.G. Radaelli for helpful discussions. This work was supported by the U.S. DOE, Office of Science under Contract. W-31-109-ENG-38.
\end{acknowledgments}

%\bibliography{CE5}

\begin{thebibliography}{22}
%\scriptsize
\expandafter\ifx\csname natexlab\endcsname\relax\def\natexlab#1{#1}\fi
\expandafter\ifx\csname bibnamefont\endcsname\relax
  \def\bibnamefont#1{#1}\fi
\expandafter\ifx\csname bibfnamefont\endcsname\relax
  \def\bibfnamefont#1{#1}\fi
\expandafter\ifx\csname citenamefont\endcsname\relax
  \def\citenamefont#1{#1}\fi
\expandafter\ifx\csname url\endcsname\relax
  \def\url#1{\texttt{#1}}\fi
\expandafter\ifx\csname urlprefix\endcsname\relax\def\urlprefix{URL }\fi
\providecommand{\bibinfo}[2]{#2}
\providecommand{\eprint}[2][]{\url{#2}}

\bibitem[{\citenamefont{Millis et~al.}(1995)\citenamefont{Millis, Littlewood,
  and Shraiman}}]{Millis}
\bibinfo{author}{\bibfnamefont{A.~J.} \bibnamefont{Millis}},
%  \bibinfo{author}{\bibfnamefont{P.~B.} \bibnamefont{Littlewood}},
%  \bibnamefont{and} \bibinfo{author}{\bibfnamefont{B.~I.}
%  \bibnamefont{Shraiman}}, 
  \bibnamefont{et~al.}, \bibinfo{journal}{Phys. Rev. Lett.}
  \textbf{\bibinfo{volume}{74}}, \bibinfo{pages}{5144} (\bibinfo{year}{1995}).

\bibitem[{\citenamefont{Roder et~al.}(1996)\citenamefont{Roder, Zang, and
  Bishop}}]{Roder}
\bibinfo{author}{\bibfnamefont{H.}~\bibnamefont{Roder}},
%  \bibinfo{author}{\bibfnamefont{J.}~\bibnamefont{Zang}}, \bibnamefont{and}
 % \bibinfo{author}{\bibfnamefont{A.~R.} \bibnamefont{Bishop}},
   \bibnamefont{et~al.}, \bibinfo{journal}{Phys. Rev. Lett.} \textbf{\bibinfo{volume}{76}},
  \bibinfo{pages}{1356} (\bibinfo{year}{1996}).

\bibitem[{\citenamefont{Murakami et~al.}(1998)\citenamefont{Murakami, Hill,
  Gibbs, Blume, Koyama, Tanaka, Kawata, Arima, Tokura, Hirota
  et~al.}}]{cjteffect}
\bibinfo{author}{\bibfnamefont{Y.}~\bibnamefont{Murakami}},
%  \bibinfo{author}{\bibfnamefont{J.~P.} \bibnamefont{Hill}},
 % \bibinfo{author}{\bibfnamefont{D.}~\bibnamefont{Gibbs}},
 % \bibinfo{author}{\bibfnamefont{M.}~\bibnamefont{Blume}},
  %\bibinfo{author}{\bibfnamefont{I.}~\bibnamefont{Koyama}},
  %\bibinfo{author}{\bibfnamefont{M.}~\bibnamefont{Tanaka}},
  %\bibinfo{author}{\bibfnamefont{H.}~\bibnamefont{Kawata}},
  %\bibinfo{author}{\bibfnamefont{T.}~\bibnamefont{Arima}},
  %\bibinfo{author}{\bibfnamefont{Y.}~\bibnamefont{Tokura}},
  %\bibinfo{author}{\bibfnamefont{K.}~\bibnamefont{Hirota}},
  \bibnamefont{et~al.}, \bibinfo{journal}{Phys. Rev. Lett.}
  \textbf{\bibinfo{volume}{81}}, \bibinfo{pages}{582} (\bibinfo{year}{1998}).

\bibitem[{\citenamefont{Goodenough}(1955)}]{goodenough}
\bibinfo{author}{\bibfnamefont{J.}~\bibnamefont{Goodenough}},
  \bibinfo{journal}{Phys. Rev.} \textbf{\bibinfo{volume}{100}},
  \bibinfo{pages}{564} (\bibinfo{year}{1955}).

\bibitem[{\citenamefont{Radaelli et~al.}(1997)\citenamefont{Radaelli, Cox,
  Marezio, and Cheong}}]{Radaelli2}
\bibinfo{author}{\bibfnamefont{P.~G.} \bibnamefont{Radaelli}},
  %\bibinfo{author}{\bibfnamefont{D.~E.} \bibnamefont{Cox}},
  %\bibinfo{author}{\bibfnamefont{M.}~\bibnamefont{Marezio}}, \bibnamefont{and}
  %\bibinfo{author}{\bibfnamefont{S.-W.} \bibnamefont{Cheong}},
    \bibnamefont{et~al.},\bibinfo{journal}{Phys. Rev. B} \textbf{\bibinfo{volume}{55}},
  \bibinfo{pages}{3015} (\bibinfo{year}{1997}).

\bibitem[{\citenamefont{Huang et~al.}(2000)\citenamefont{Huang, Lynn, Erwin,
  Santoro, Dender, Smolyaninova, Ghosh, and Greene}}]{Huang}
\bibinfo{author}{\bibfnamefont{Q.}~\bibnamefont{Huang}},
  %\bibinfo{author}{\bibfnamefont{J.~W.} \bibnamefont{Lynn}},
  %\bibinfo{author}{\bibfnamefont{R.~W.} \bibnamefont{Erwin}},
  %\bibinfo{author}{\bibfnamefont{A.}~\bibnamefont{Santoro}},
  %\bibinfo{author}{\bibfnamefont{D.~C.} \bibnamefont{Dender}},
 % \bibinfo{author}{\bibfnamefont{V.~N.} \bibnamefont{Smolyaninova}},
  %\bibinfo{author}{\bibfnamefont{K.}~\bibnamefont{Ghosh}}, \bibnamefont{and}
  %\bibinfo{author}{\bibfnamefont{R.~L.} \bibnamefont{Greene}},
    \bibnamefont{et~al.},\bibinfo{journal}{Phys. Rev. B} \textbf{\bibinfo{volume}{61}},
  \bibinfo{pages}{8895} (\bibinfo{year}{2000}).

\bibitem[{\citenamefont{Argyriou et~al.}(2000)\citenamefont{Argyriou, Bordallo,
  Campbell, Cheetham, Cox, Gardner, Hanif, dos Santos, and
  Strouse}}]{Argyriou2}
\bibinfo{author}{\bibfnamefont{D.~N.} \bibnamefont{Argyriou}},
  %\bibinfo{author}{\bibfnamefont{H.~N.} \bibnamefont{Bordallo}},
  %\bibinfo{author}{\bibfnamefont{B.~J.} \bibnamefont{Campbell}},
  %\bibinfo{author}{\bibfnamefont{A.~K.} \bibnamefont{Cheetham}},
  %\bibinfo{author}{\bibfnamefont{D.~E.} \bibnamefont{Cox}},
  %\bibinfo{author}{\bibfnamefont{J.~S.} \bibnamefont{Gardner}},
  %\bibinfo{author}{\bibfnamefont{K.}~\bibnamefont{Hanif}},
  %\bibinfo{author}{\bibfnamefont{A.}~\bibnamefont{dos Santos}},
  %\bibnamefont{and} \bibinfo{author}{\bibfnamefont{G.~F.}\bibnamefont{Strouse}}, 
  \bibnamefont{et~al.},\bibinfo{journal}{Phys. Rev. B}
  \textbf{\bibinfo{volume}{61}}, \bibinfo{pages}{15269} (\bibinfo{year}{2000}).

\bibitem[{\citenamefont{Vasiliu-Doloc et~al.}(1999)\citenamefont{Vasiliu-Doloc,
  Rosenkranz, Osborn, Sinha, Lynn, Mesot, Seeck, Preosti, Fedro, and
  Mitchell}}]{Doloc}
\bibinfo{author}{\bibfnamefont{L.}~\bibnamefont{Vasiliu-Doloc}},
  %\bibinfo{author}{\bibfnamefont{S.}~\bibnamefont{Rosenkranz}},
  %\bibinfo{author}{\bibfnamefont{R.}~\bibnamefont{Osborn}},
  %\bibinfo{author}{\bibfnamefont{S.~K.} \bibnamefont{Sinha}},
  %\bibinfo{author}{\bibfnamefont{J.~W.} \bibnamefont{Lynn}},
  %\bibinfo{author}{\bibfnamefont{J.}~\bibnamefont{Mesot}},
  %\bibinfo{author}{\bibfnamefont{O.~H.} \bibnamefont{Seeck}},
  %\bibinfo{author}{\bibfnamefont{G.}~\bibnamefont{Preosti}},
  %\bibinfo{author}{\bibfnamefont{A.~J.} \bibnamefont{Fedro}}, \bibnamefont{and}
  %\bibinfo{author}{\bibfnamefont{J.~F.} \bibnamefont{Mitchell}},
    \bibnamefont{et~al.},\bibinfo{journal}{Phys. Rev. Lett.} \textbf{\bibinfo{volume}{83}},
  \bibinfo{pages}{4393} (\bibinfo{year}{1999}).

\bibitem[{\citenamefont{Argyriou et~al.}(1999)\citenamefont{Argyriou, Bordallo,
  Mitchell, Jorgensen, and Strouse}}]{Argyriou}
\bibinfo{author}{\bibfnamefont{D.~N.} \bibnamefont{Argyriou}},
  %\bibinfo{author}{\bibfnamefont{H.~N.} \bibnamefont{Bordallo}},
  %\bibinfo{author}{\bibfnamefont{J.~F.} \bibnamefont{Mitchell}},
  %\bibinfo{author}{\bibfnamefont{J.~D.} \bibnamefont{Jorgensen}},
  %\bibnamefont{and} \bibinfo{author}{\bibfnamefont{G.~F.}
  %\bibnamefont{Strouse}}, 
  \bibnamefont{et~al.},\bibinfo{journal}{Phys. Rev. B}
  \textbf{\bibinfo{volume}{60}}, \bibinfo{pages}{6200} (\bibinfo{year}{1999}).

\bibitem[{\citenamefont{Shimomura et~al.}(1999)\citenamefont{Shimomura,
  Wakabayashi, Kuwahara, and Tokura}}]{Shimo}
\bibinfo{author}{\bibfnamefont{S.}~\bibnamefont{Shimomura}},
  %\bibinfo{author}{\bibfnamefont{N.}~\bibnamefont{Wakabayashi}},
  %\bibinfo{author}{\bibfnamefont{H.}~\bibnamefont{Kuwahara}}, \bibnamefont{and}
  %\bibinfo{author}{\bibfnamefont{Y.}~\bibnamefont{Tokura}},
    \bibnamefont{et~al.},\bibinfo{journal}{Phys. Rev. Lett.} \textbf{\bibinfo{volume}{83}},
  \bibinfo{pages}{4389} (\bibinfo{year}{1999}).

\bibitem[{\citenamefont{Adams et~al.}(2000)\citenamefont{Adams, Lynn,
  Mukovskii, Arsenov, and Shulyatev}}]{Adams}
\bibinfo{author}{\bibfnamefont{C.~P.} \bibnamefont{Adams}},
  %\bibinfo{author}{\bibfnamefont{J.~W.} \bibnamefont{Lynn}},
  %\bibinfo{author}{\bibfnamefont{Y.~M.} \bibnamefont{Mukovskii}},
  %\bibinfo{author}{\bibfnamefont{A.~A.} \bibnamefont{Arsenov}},
  %\bibnamefont{and} \bibinfo{author}{\bibfnamefont{D.~A.}
  %\bibnamefont{Shulyatev}}, 
  \bibnamefont{et~al.},\bibinfo{journal}{Phys. Rev. Lett.}
  \textbf{\bibinfo{volume}{85}}, \bibinfo{pages}{3954} (\bibinfo{year}{2000}).

\bibitem[{\citenamefont{Campbell et~al.}(2001)\citenamefont{Campbell, Osborn,
  Argyriou, Vasiliu-Doloc, Mitchell, Sinha, Ruett, Ling, Islam, and
  W.}}]{campbell}
\bibinfo{author}{\bibfnamefont{B.~J.} \bibnamefont{Campbell}},
  %\bibinfo{author}{\bibfnamefont{R.}~\bibnamefont{Osborn}},
  %\bibinfo{author}{\bibfnamefont{D.~N.} \bibnamefont{Argyriou}},
  %\bibinfo{author}{\bibfnamefont{L.}~\bibnamefont{Vasiliu-Doloc}},
  %\bibinfo{author}{\bibfnamefont{J.~F.} \bibnamefont{Mitchell}},
  %\bibinfo{author}{\bibfnamefont{S.~K.} \bibnamefont{Sinha}},
  %\bibinfo{author}{\bibfnamefont{U.}~\bibnamefont{Ruett}},
  %\bibinfo{author}{\bibfnamefont{C.~D.} \bibnamefont{Ling}},
  %\bibinfo{author}{\bibfnamefont{Z.}~\bibnamefont{Islam}}, \bibnamefont{and}
  %\bibinfo{author}{\bibfnamefont{L.~J.} \bibnamefont{W.}},
    \bibnamefont{et~al.},\bibinfo{journal}{Phys. Rev. B} \textbf{\bibinfo{volume}{65}},
  \bibinfo{pages}{014427} (\bibinfo{year}{2001}).

\bibitem[{\citenamefont{Dai et~al.}(2000)\citenamefont{Dai, Fernandez-Baca,
  Wakabayashi, Plummer, Tomioka, and Tokura}}]{Dai}
\bibinfo{author}{\bibfnamefont{P.~C.} \bibnamefont{Dai}},
  %\bibinfo{author}{\bibfnamefont{J.~A.} \bibnamefont{Fernandez-Baca}},
  %\bibinfo{author}{\bibfnamefont{N.}~\bibnamefont{Wakabayashi}},
  %\bibinfo{author}{\bibfnamefont{E.~W.} \bibnamefont{Plummer}},
  %\bibinfo{author}{\bibfnamefont{Y.}~\bibnamefont{Tomioka}}, \bibnamefont{and}
  %\bibinfo{author}{\bibfnamefont{Y.}~\bibnamefont{Tokura}},
    \bibnamefont{et~al.},\bibinfo{journal}{Phys. Rev. Lett.} \textbf{\bibinfo{volume}{85}},
  \bibinfo{pages}{2553} (\bibinfo{year}{2000}).

\bibitem[{\citenamefont{Mitchell et~al.}(1997)\citenamefont{Mitchell, Argyriou,
  Jorgensen, Hinks, Potter, and Bader}}]{Mitchell}
\bibinfo{author}{\bibfnamefont{J.}~\bibnamefont{Mitchell}},
  %\bibinfo{author}{\bibfnamefont{D.}~\bibnamefont{Argyriou}},
  %\bibinfo{author}{\bibfnamefont{J.}~\bibnamefont{Jorgensen}},
  %\bibinfo{author}{\bibfnamefont{D.}~\bibnamefont{Hinks}},
  %\bibinfo{author}{\bibfnamefont{C.}~\bibnamefont{Potter}}, \bibnamefont{and}
  %\bibinfo{author}{\bibfnamefont{S.}~\bibnamefont{Bader}},
    \bibnamefont{et~al.},\bibinfo{journal}{Phys. Rev. B} \textbf{\bibinfo{volume}{55}},
  \bibinfo{pages}{63} (\bibinfo{year}{1997}).

\bibitem[{\citenamefont{Gardner et~al.}(1999)\citenamefont{Gardner, Gaulin,
  Lee, Broholm, Raju, and Greedan}}]{Gardner}
\bibinfo{author}{\bibfnamefont{J.}~\bibnamefont{Gardner}},
  %\bibinfo{author}{\bibfnamefont{B.}~\bibnamefont{Gaulin}},
  %\bibinfo{author}{\bibfnamefont{S.-H.} \bibnamefont{Lee}},
  %\bibinfo{author}{\bibfnamefont{C.}~\bibnamefont{Broholm}},
  %\bibinfo{author}{\bibfnamefont{N.}~\bibnamefont{Raju}}, \bibnamefont{and}
  %\bibinfo{author}{\bibfnamefont{J.}~\bibnamefont{Greedan}},
    \bibnamefont{et~al.},\bibinfo{journal}{Phys. Rev. Lett.} \textbf{\bibinfo{volume}{83}},
  \bibinfo{pages}{211} (\bibinfo{year}{1999}).

\bibitem[{\citenamefont{Gordon et~al.}(1999)\citenamefont{Gordon, Bader,
  Mitchell, Osborn, and Rosenkranz}}]{Gordon}
\bibinfo{author}{\bibfnamefont{J.}~\bibnamefont{Gordon}},
  %\bibinfo{author}{\bibfnamefont{S.}~\bibnamefont{Bader}},
  %\bibinfo{author}{\bibfnamefont{J.}~\bibnamefont{Mitchell}},
  %\bibinfo{author}{\bibfnamefont{R.}~\bibnamefont{Osborn}}, \bibnamefont{and}
  %\bibinfo{author}{\bibfnamefont{S.}~\bibnamefont{Rosenkranz}},
    \bibnamefont{et~al.},\bibinfo{journal}{Phys. Rev. B} \textbf{\bibinfo{volume}{60}},
  \bibinfo{pages}{6258} (\bibinfo{year}{1999}).

\bibitem[{\citenamefont{McQueeney et~al.}(1999)\citenamefont{McQueeney, Petrov,
  Egami, Yethiraj, Shirane, and Endoh}}]{McQueeney}
\bibinfo{author}{\bibfnamefont{R.}~\bibnamefont{McQueeney}},
  %\bibinfo{author}{\bibfnamefont{Y.}~\bibnamefont{Petrov}},
  %\bibinfo{author}{\bibfnamefont{T.}~\bibnamefont{Egami}},
  %\bibinfo{author}{\bibfnamefont{M.}~\bibnamefont{Yethiraj}},
  %\bibinfo{author}{\bibfnamefont{G.}~\bibnamefont{Shirane}}, \bibnamefont{and}
  %\bibinfo{author}{\bibfnamefont{Y.}~\bibnamefont{Endoh}},
    \bibnamefont{et~al.},\bibinfo{journal}{Phys. Rev. Lett.} \textbf{\bibinfo{volume}{82}},
  \bibinfo{pages}{628} (\bibinfo{year}{1999}).

\bibitem[{\citenamefont{Zhang et~al.}(2001)\citenamefont{Zhang, Dai,
  Fernandez-Baca, Plummer, Tomioka, and Tokura}}]{Zhang}
\bibinfo{author}{\bibfnamefont{J.~D.} \bibnamefont{Zhang}},
  %\bibinfo{author}{\bibfnamefont{P.~C.} \bibnamefont{Dai}},
  %\bibinfo{author}{\bibfnamefont{J.~A.} \bibnamefont{Fernandez-Baca}},
  %\bibinfo{author}{\bibfnamefont{E.~W.} \bibnamefont{Plummer}},
  %\bibinfo{author}{\bibfnamefont{Y.}~\bibnamefont{Tomioka}}, \bibnamefont{and}
  %\bibinfo{author}{\bibfnamefont{Y.}~\bibnamefont{Tokura}},
    \bibnamefont{et~al.},\bibinfo{journal}{Phys. Rev. Lett.} \textbf{\bibinfo{volume}{86}},
  \bibinfo{pages}{3823} (\bibinfo{year}{2001}).

\bibitem[{\citenamefont{Burgy et~al.}(2001)\citenamefont{Burgy, Mayr,
  Martin-Mayor, Moreo, and Dagotto}}]{burgy}
\bibinfo{author}{\bibfnamefont{J.}~\bibnamefont{Burgy}},
  %\bibinfo{author}{\bibfnamefont{M.}~\bibnamefont{Mayr}},
  %\bibinfo{author}{\bibfnamefont{V.}~\bibnamefont{Martin-Mayor}},
  %\bibinfo{author}{\bibfnamefont{A.}~\bibnamefont{Moreo}}, \bibnamefont{and}
  %\bibinfo{author}{\bibfnamefont{E.}~\bibnamefont{Dagotto}},
    \bibnamefont{et~al.},\bibinfo{journal}{Phys. Rev. Lett.} \textbf{\bibinfo{volume}{87}},
  \bibinfo{pages}{277202} (\bibinfo{year}{2001}).

\bibitem[{\citenamefont{Tranquada et~al.}(1997)\citenamefont{Tranquada, Axe,
  Ichikawa, Moodenbaugh, Nakamura, and Uchida}}]{Tranquada}
\bibinfo{author}{\bibfnamefont{J.}~\bibnamefont{Tranquada}},
  %\bibinfo{author}{\bibfnamefont{J.}~\bibnamefont{Axe}},
  %\bibinfo{author}{\bibfnamefont{N.}~\bibnamefont{Ichikawa}},
  %\bibinfo{author}{\bibfnamefont{A.}~\bibnamefont{Moodenbaugh}},
  %\bibinfo{author}{\bibfnamefont{Y.}~\bibnamefont{Nakamura}}, \bibnamefont{and}
  %\bibinfo{author}{\bibfnamefont{S.}~\bibnamefont{Uchida}},
    \bibnamefont{et~al.},\bibinfo{journal}{Phys. Rev. Lett.} \textbf{\bibinfo{volume}{78}},
  \bibinfo{pages}{338} (\bibinfo{year}{1997}).

\bibitem[{\citenamefont{Lanzara et~al.}(2001)\citenamefont{Lanzara, Bogdanov,
  Zhou, Kellar, Geng, Lu, Yoshida, Eisaki, Fujimori, Kishio et~al.}}]{lanzara}
\bibinfo{author}{\bibfnamefont{A.}~\bibnamefont{Lanzara}},
  %\bibinfo{author}{\bibfnamefont{P.~V.} \bibnamefont{Bogdanov}},
  %\bibinfo{author}{\bibfnamefont{X.~J.} \bibnamefont{Zhou}},
  %\bibinfo{author}{\bibfnamefont{S.~A.} \bibnamefont{Kellar}},
  %\bibinfo{author}{\bibfnamefont{D.~L.} \bibnamefont{Geng}},
  %\bibinfo{author}{\bibfnamefont{E.~D.} \bibnamefont{Lu}},
  %\bibinfo{author}{\bibfnamefont{T.}~\bibnamefont{Yoshida}},
  %\bibinfo{author}{\bibfnamefont{H.}~\bibnamefont{Eisaki}},
  %\bibinfo{author}{\bibfnamefont{A.}~\bibnamefont{Fujimori}},
  %\bibinfo{author}{\bibfnamefont{K.}~\bibnamefont{Kishio}},
\bibnamefont{et~al.}, \bibinfo{journal}{Nature}
  \textbf{\bibinfo{volume}{412}} (\bibinfo{year}{2001}).

\bibitem[{\citenamefont{Millis}(1996)}]{Millis2}
\bibinfo{author}{\bibfnamefont{A.~J.} \bibnamefont{Millis}},
  \bibinfo{journal}{Phys. Rev. B} \textbf{\bibinfo{volume}{53}},
  \bibinfo{pages}{8434} (\bibinfo{year}{1996}).

\end{thebibliography}

%\subsubsection{}

% If in two-column mode, this environment will change to single-column
% format so that long equations can be displayed. Use
%\begin{widetext}
%\end{widetext}

% figures should be put into the text as floats.

\end{document}